\documentstyle[prl,epsfig,twocolumn,aps]{revtex}
\begin{document} 
\twocolumn[
\draft
\title{An Accurate Determination of the Exchange Constant in $\rm Sr_2CuO_3$
from Recent Theoretical Results}
\author{Sebastian Eggert}
\address{Theoretical Physics, Chalmers University of Technology
 and G\"oteborg University, 41296 Gothenburg, Sweden, (eggert@fy.chalmers.se)}
\maketitle 
\date{\today} 
\begin{abstract}
\widetext\leftskip=0.10753\textwidth \rightskip\leftskip
Data from susceptibility measurements\cite{dupont} on $\rm Sr_2CuO_3$ are 
compared with recent theoretical predictions\cite{eggert} for the magnetic
susceptibility of the antiferromagnetic spin-1/2 Heisenberg chain.
The experimental data fully confirms the theoretical predictions
and in turn we establish that $\rm Sr_2CuO_3$
behaves almost perfectly like a one-dimensional antiferromagnet with   
an exchange coupling of  $J = 1700^{+150}_{-100}K$.
\end{abstract}] \narrowtext
The Hamiltonian for the antiferromagnetic spin-1/2 Heisenberg chain
\begin{equation} H = J \sum_i \vec{S}_i \cdot \vec{S}_{i+1} 
\label{ham} \end{equation} 
has been a popular and very well studied model in theoretical 
physics for a very long time.   It was not until recently, however,
that the bulk susceptibility per site 
\begin{equation} \chi(T) \equiv \frac{g^2 \mu_B^2}{ k_B T} 
\sum_i < S^z_0 S^z_i >_{{}_T}^{} 
\end{equation}
was calculated for the full
temperature range by combining analytical arguments from field theory and 
numerical results from the algebraic Bethe ansatz equations\cite{eggert}. 
That work has to be viewed in relation to the pioneering results
of Bonner and Fisher\cite{bonner}, who calculated the susceptibility
of finite chains numerically more than 30 years ago. 
The results of Bonner and Fisher 
correctly predicted a broad maximum in the susceptibility $\chi(T)$ 
 which was very useful for determining the characteristics of quasi
one-dimensional materials with moderate values of $J$\cite{studies,CPC}.  
Attempts to extrapolate the Bonner and Fisher 
data to lower temperatures\cite{extrapol}, however, turned out to
yield incorrect results.  

The most surprising result of reference \onlinecite{eggert} is the 
prediction of a divergent slope of $\chi(T)$ at $T=0$ together with
an inflection point at $T \approx 0.087 J$.
At low temperatures the deviation of the extrapolated 
Bonner and Fisher curve from this exotic behavior is 
quite significant as shown in figure (\ref{susc}).  The value
at zero temperature is $\chi(0) = 1/J \pi^2$ (we set the
gyromagnetic ratio times the Bohr magneton $g \mu_B$ as well as the
Boltzmann constant $k_B$ to unity, so that the susceptibility is
measured in units of $1/J$ and the temperature is measured in units of $J$).

\begin{figure}
\begin{center}
\mbox{\epsfig{file=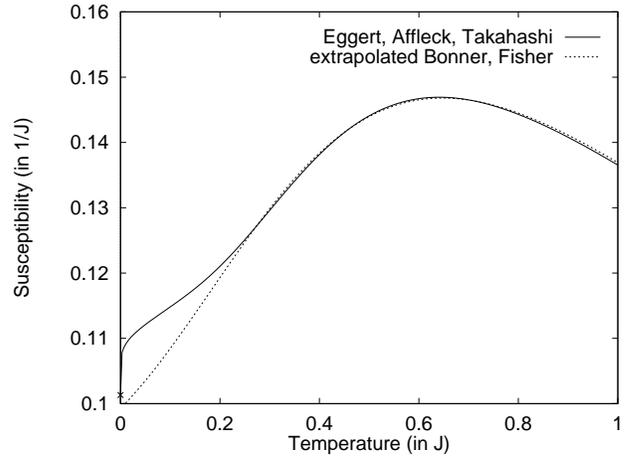,width=3.375in}}
\end{center}
\caption{The susceptibility of the spin-1/2 chain according to
recent calculations\protect{\cite{eggert}} compared to the extrapolated
Bonner and Fisher curve (from \protect{\onlinecite{extrapol}}).}
\label{susc}
\end{figure}
We will now attempt to present experimental evidence for the
surprising behavior of the susceptibility at low temperatures.
The one-dimensional characteristics of all experimental systems
break down at some finite temperature  (because of a spin-Peierls
transition or three dimensional ordering),
but nonetheless we  expect that the best
quasi one-dimensional materials should show an inflection point 
at low temperatures and a significant deviation from the 
extrapolated Bonner and Fisher curve.

The material $\rm Sr_2CuO_3$ is believed to have the best one-dimensional
characteristics of an antiferromagnet reported so far\cite{dupont}. 
In particular, it is one of the few materials which will be able to exhibit
the difference between the two curves in figure (\ref{susc}).  (In fact,
it is probably the only known material to have a low enough transition
temperature 
besides $\rm CuCl_2 \cdot 2NC_5H_5$, for which a clear deviation at low 
temperatures from the Bonner and Fisher curve was first noticed in the
early 70's\cite{CPC}.) $\rm Sr_2CuO_3$ is very much
of current interest since it is directly related to high temperature
superconductors and $\rm Sr_2CuO_{3.1}$ was reported to exhibit
high temperature superconductivity ($T_c \approx 70K$)
under high presure\cite{SC}.
The exchange constant $J$ in equation (\ref{ham}) is expected to
be roughly the same as the Cu-Cu super-exchange interaction in the 
layered cuprates since the Cu-Cu distances are comparable.

A number of susceptibility measurements have been performed on this
material \cite{other,neelorder} 
and rather good results are available from recent
measurements on carefully prepared, high quality samples of $\rm Sr_2CuO_3$
by T. Ami et.al.\cite{dupont}. Their data was 
analyzed under the assumption that an extrapolated Bonner and Fisher
curve yielded good results also for lower temperatures. They reported a
general agreement with the Bonner and Fisher curve and an exchange
constant of $J = 2600^{+200}_{-400}K$.

We took the identical experimental data and performed a fit 
according to the newly
available data from the numerical Bethe ansatz calculations
of reference \onlinecite{eggert} but using otherwise identical
assumptions.   Namely, after subtracting the core diamagnetism 
(from reference \onlinecite{dia}), we fitted the 
total susceptibility $\chi^{tot}$
assuming a constant term from Van Vleck paramagnetism $\chi^{VV}$,
a Curie-Weiss term per impurity
$\chi^{CW}(T) = \frac{g^2 \mu_B^2 S (S+1)}{3 k_B (T - \Theta)}$,
and the spin chain part $\chi(T)$ from reference \onlinecite{eggert}:  
\begin{equation}
\chi^{tot}(T) =  \chi(T) + \rho \chi^{CW}(T) + \chi^{VV},
\end{equation} where $\rho$ is the impurity density 
(assuming nearly isolated finite length spin-chains with an 
odd number of spins).

The result of our fit is shown in figure (\ref{fit}) which yielded
a dramatically different estimate for the exchange constant
\begin{equation} J = 1700^{+150}_{-100}K \end{equation}
compared to $J = 2600K$ in reference \onlinecite{dupont}
(taking into account their different definition of $J$ by a factor of
two).
Other parameters in our fit are $\theta \approx -4.49K$, a Van Vleck 
susceptibility of $\chi^{VV} \approx 2.55 \times 10^{-5} \rm ccm/mole$
and an impurity density of $\rho \approx 0.16 \%$. 

\begin{figure}
\begin{center}
\mbox{\epsfig{file=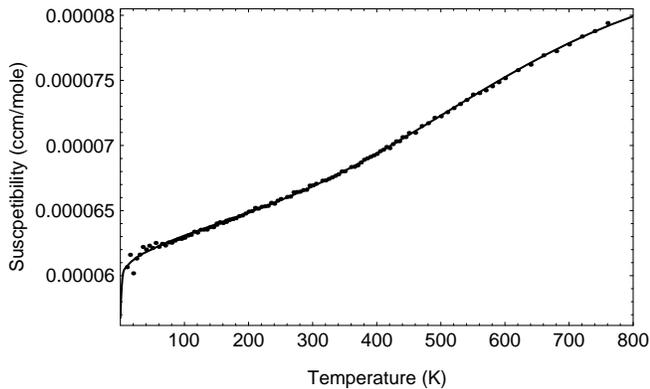,width=3.375in}}
\end{center}
\caption{The fit of the magnetic susceptibility according
to the results of reference \protect{\onlinecite{eggert}}
with $J = 1700K$.  The background susceptibility and the
Curie-Weiss term are subtracted.} \label{fit}
\end{figure}

Our greatly different estimate of $J$, however,
is based on a much better fit of the experimental data. The 
deviation of the experimental data from the least squares
fit is plotted in figure 
(\ref{quality}) for two different cases: \begin{itemize}
\item{\bf (A)} our fit according to reference \onlinecite{eggert} ($J = 1700K$)
\item{\bf(B)}  the fit according to the extrapolated Bonner and Fisher curve
($J=2614K$)
\end{itemize} (fit {\bf (B)} was taken directly from reference 
\onlinecite{dupont} which in turn was based on references \onlinecite{bonner}
and \onlinecite{extrapol}).  

\begin{figure}
\begin{center}
\mbox{\epsfig{file=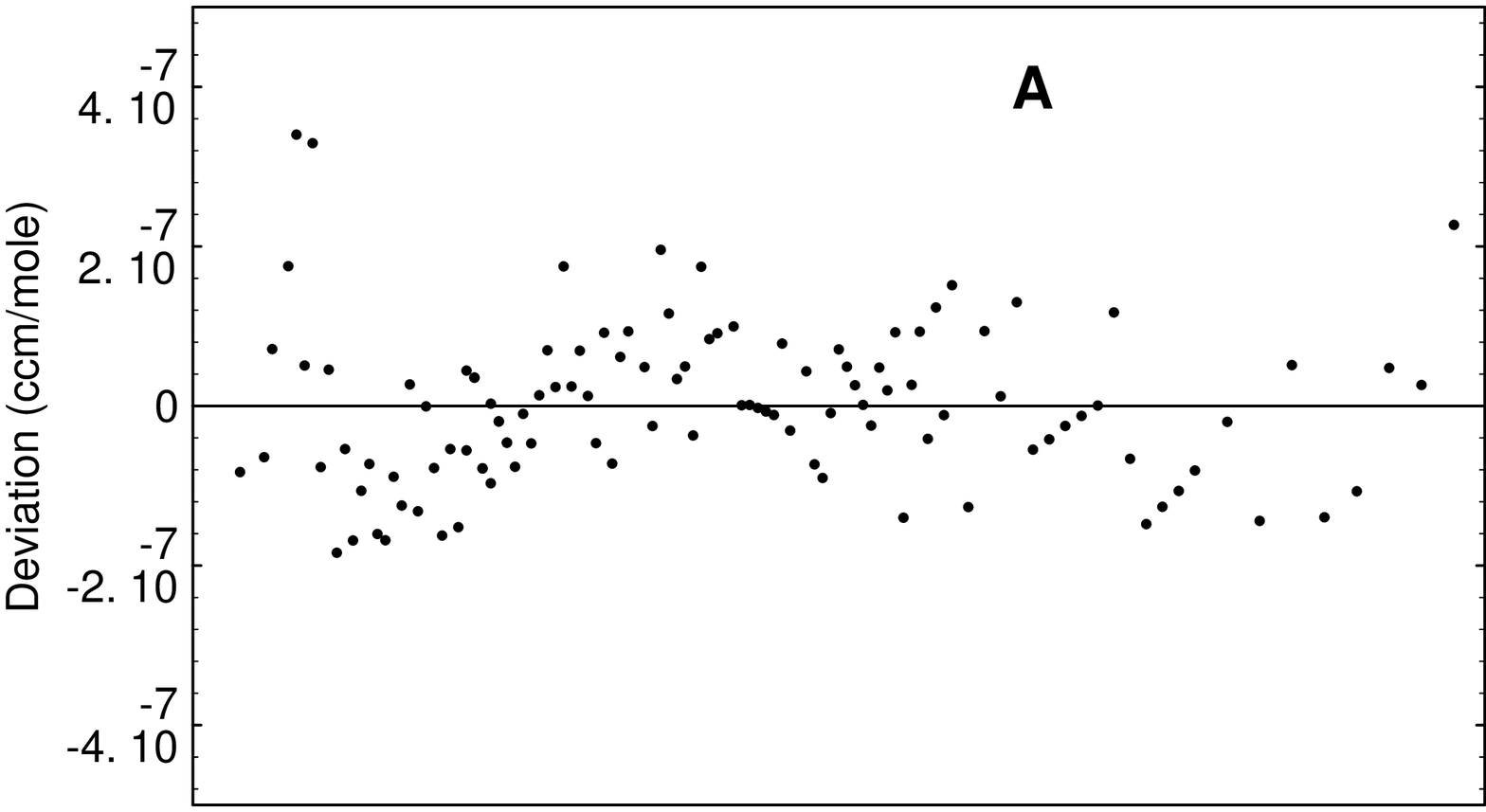,width=3.375in}}
\mbox{\epsfig{file=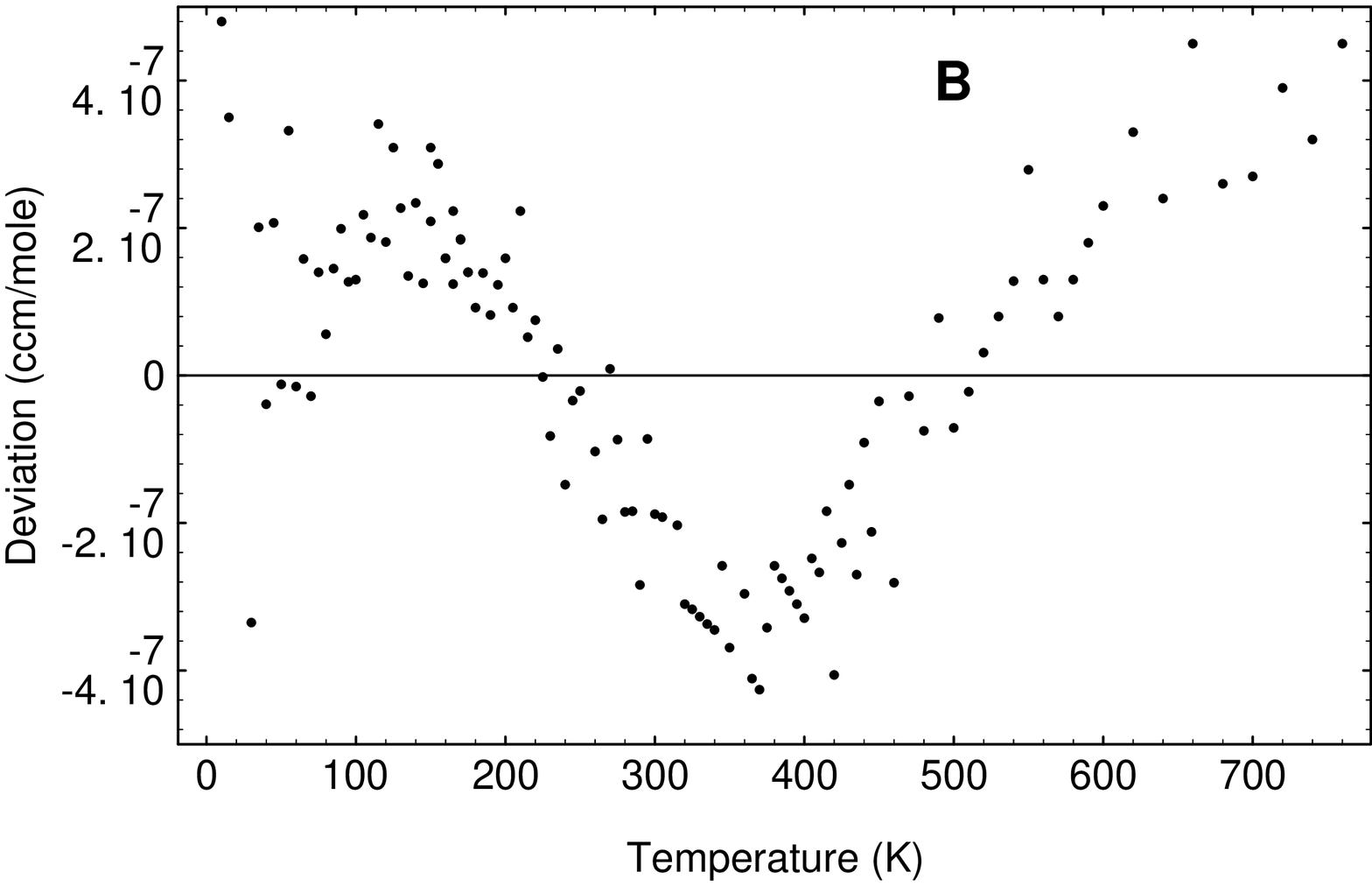,width=3.375in}}
\end{center}
\caption{ The deviation of the experimental data from the least squares fit
using the results of Eggert et. al. \protect{\cite{eggert}} {\bf(A)} and
using the extrapolated Bonner and Fisher result
\protect{\cite{dupont}} {\bf (B)}.} \label{quality}
\end{figure}
We can see that the fit {\bf(B) }to the extrapolated Bonner and Fisher
curve contains a systematic deviation which cannot be explained by
experimental error. Our new fit {\bf(A)}, however, 
is fully within the statistical 
fluctuations of the experimental data.  Both fits have random
deviations at very low temperatures that are larger than the scale of the
graph (\ref{quality}), which may be due to larger experimental error in that
region.  In any case we cannot expect that a Curie-Weiss term is  
fully adequate to describe impurity effects 
in that temperature region. However, this has little effect on the
fit over the full temperature range or on the estimate of $J$.

We take the extremely good fit  
as strong evidence that reference \onlinecite{eggert} predicted
the susceptibility of the one-dimensional Heisenberg model correctly.
Moreover, the quality of the fit establishes that $\rm Sr_2CuO_3$ 
is very well described by the model in equation (\ref{ham}) over a large
temperature range.
There has been no report of a spin-Peierls transition in this
material and the three dimensional ordering temperature was reported
to be $T_N \approx 5K$ from $\mu SR$ experiments\cite{neelorder}.
This makes $\rm Sr_2CuO_3$ the  
antiferromagnet with the best quasi one-dimensional characteristics
reported so far $J/T_N \approx 300$.

 In conclusion we have presented strong experimental confirmation for the 
predicted exotic temperature dependence of the Heisenberg chain 
susceptibility at low temperatures\cite{eggert}. 
The material $\rm Sr_2CuO_3$ has been established as  
a highly one-dimensional antiferromagnet with a much improved estimate
of the exchange constant $J = 1700^{+150}_{-100}K $ which compares 
rather well with the values of the exchange interaction in the 
layered cuprates ($\approx 1480\pm80K$\cite{cuprateJ}).  An independent 
experimental check of the exchange constant $J$ in $\rm Sr_2CuO_3$
would certainly be desirable.
\begin{acknowledgements}
The author  
expresses his gratitude to Micheal Crawford of Du Pont and
David Johnston of Ames Laboratory for supplying the
experimental data and helpful comments. 
I would also like to thank Ian Affleck, Henrik Johannesson, Ann Mattsson,
Stefan Rommer, Fabian Wenger, and Stellan \"Ostlund
for helpful discussions.  The work was supported in part 
by the Swedish Natural Science Research Council.
\end{acknowledgements}

\end{document}